\documentstyle{article}
\begin{document}
\title{\Large\bf Transport and magnetic properties in ferromagnetic
manganese-oxide thin films \\}
\author{\large Liang-Jian Zou \\
{\it  Institute of Solid State Physics, Academia Sinica,
      P.O.Box 1129, Hefei 230031, China}  \\
\large X. G. Gong \\
{\it  CCAST (World Laboratory) P.O.Box 8730, Beijing, 100080 } \\
{\it  Institute of Solid State Physics, Academia Sinica,
      P.O.Box 1129, Hefei 230031, China}  \\
\large Qing-Qi Zheng \\
{\it  Institute of Solid State Physics, Academia Sinica,
      P.O.Box 1129, Hefei 230031, China} \\
\large C. Y. Pan \\
{\it Center of Magnetism in Information Technology, Department of Physics,}\\
{\it Utah State University, Logan,UT 84322-4415 } \\ }

\date{ }
\maketitle
\large
\begin{center}
{\bf Abstract}
\end{center}
   The transport and magnetic properties in ferromagnetic manganese-oxide
thin films are studied based on the model of the coupling between the
mobile d-electrons and the core spins in Mn ions. The spontaneous
magnetization and the resistivity are obtained for various magnetic
fields and temperature. The resistivity in absence of magnetic field
and the magnetoresistance exhibit maxima near the Curie temperature, the
applied magnetic field moves the position of the resistivity peak to
high temperature and suppresses the peak value, which agree with the
experimental results. The Hall resistivity is predicted to exhibit
maximum near the Curie point. The pressure effect of the magnetoresistance
can also be explained qualitatively in this mechanism. The colossal
magnetoresistance in ferromagnetic manganese-oxide thin films is
attributed to the spin-correlation fluctuation scattering and the
low dimensional effect.

\vspace{1cm}

\newpage
\noindent {\bf I. INTRODUCTION} \\

    In perovskite-type transition-metal oxides, the strong correlated electron
systems exhibit many remarkable phenomena. One example of such systems is the
high temperature superconductivity in cuprate-oxides La$_{2-x}$R$_{x}$CuO$_{4}$
and YBa$_{2}$Cu$_{3}$O$_{7-\delta}$. Recently very large negative
magnetoresistance (MR), so called colossal magnetoresistance (CMR), has been
observed in perovskite-like La$_{1-x}$R$_{x}$MnO$_{3}$ and
Nd$_{1-x}$R$_{x}$MnO$_{3}$ ( R =Ba, Sr, Ca, Pb, etc. )  [ 1 - 4 ] ferromagnetic
metallic thin films. Especially in epitaxial La$_{1-x}$Ca$_{x}$MnO$_{3}$ and
Nd$_{1-x}$Sr$_{x}$MnO$_{3}$ thin films, the MR ratios under an applied magnetic
field, $(R_{B}-R_{0})/R_{B}$ (B=6 T), are observed more than 10$^{5}$${\%}$
at low
temperature, which are much larger than those in ferromagnetic/nonmagnetic
metallic multilayers, the resistivity decreases from a typical insulate or
semiconductive behavior to metallic behavior. More recently a giant
magnetoresistance ratio is also observed to be in excess of 10$^{4}$${\%}$
 for polycrystalline La-Y-Ca-Mn-O compounds [5]. Such
CMR change provides promising applications such as in magnetic
sensor, in magnetic recording media, etc. . It also presents another
possible MR mechanism which is very different from that in magnetic multilayer.

 Similar to the high temperature superconductive material La$_{2}$CuO$_{4}$, the
undoped La$MnO_{3}$ is an antiferromagnetic [6] or a ferromagnetic [7]
insulator. X-ray photoemission spectroscopic, and ultraviolet photoemission
spectroscopic experiments on pure LaMnO$_{3}$ [8] show that the on-site
Coulomb energy of Mn 3d states is U$_{dd}$=4.0 $eV$, the hybridization
energy between Mn-O is $\Delta$=3.8 $eV$, which indicates that
La-R-Mn-O compounds are mediated-strength correlated electron
 systems. After substitution of some
trivalent La ions by Ca, Sr, Pb, Ba or other divalent ions,
a fraction of trivalent manganese ions coexist with tetravalent manganese
ions in the system, the valence fluctuation of Mn$^{+3}$ and Mn$^{+4}$
is thus assumed to be important and may give rise to the hopping
conductivity and other transport properties of these systems [9].
After doped by divalent elements R in these compounds,
La$_{2-x}$R$_{x}$CuO$_{4}$ and La$_{1-x}$R$_{x}$MnO$_{3}$
become metallic for heavy doping.
 La$_{2-x}$R$_{x}$CuO$_{4}$ may lose its resistivity below the
superconductive transition temperature, while La$_{1-x}$R$_{x}$MnO$_{3}$
exhibits giant magnetoresistance around
the Curie temperature. The superconductivity in La$_{2-x}$R$_{x}$CuO$_{4}$ and
the CMR effect in La$_{1-x}$R$_{x}$MnO$_{3}$ are supposed to be related to
the magnetic interaction.

  The mechanism of CMR has been studied by some researchers. Several mechanisms
have been suggested to explain these more than a thousandfold change of
resistivity observed in these compounds, such as
 the double-exchange and magnetic polaron
hopping [1-4], the spin spiral states [10], the Jahn-Teller effect [11] and the
spin disorder scattering [12]. Some researchers [1-5,13,14] suggested
that the magnetic polaron transfer mechanism may play a role
for the CMR effect in these
systems. As in La$_{1-x}$R$_{x}$MnO$_{3}$ compounds, the
semiconductive behavior in conduction and the negative temperature coefficient
of the resistivity above T$_{c}$ seem to support the thermal hopping of
the magnetic polarons [2,13], however, the quantitative calculation shows that
there exists considerable discrepancy between experimental and theoretical
results for the resistivity in magnetic field [11]. Besides,it seems that
magnetic polaron is hardly to form for heavy doped La$_{1-x}$R$_{x}$MnO$_{3}$
and Nd$_{1-x}$R$_{x}$MnO$_{3}$ systems (x $\ >$ 0.3). Another suggestion
is that the CMR is related to the spin disorder scattering mechanism due
to the field-induced change in the canting angle of manganese spins [12].
A strong external magnetic field will induce the spin canting of
Mn ion. However, as indicated by the experiments [3,4], the
magnetic field about 1 ${\it Tesla}$ can saturate the magnetization,
further increasing magnetic field does not change the spin configuration
significantly,
so the spin-disorder scattering shouldn't change the resistivity by several
orders in magnitude. Therefore the spin disorder scattering may be not
the main mechanism of the CMR in these systems.

      As seen from the above discussion, these mechanisms did not give a
satisfactory explain regarding the CMR
 behavior under strong magnetic field over all the temperature
ranges.  In this paper, we present another possible mechanism of
the CMR in these ferromagnetic La$_{1-x}$R$_{x}$MnO$_{3}$ and
Nd$_{1-x}$R$_{x}$MnO$_{3}$ thin films. This paper is organized as follows,
the proposed model and formalism
is described in Sec.II., Sec.III gives the results and
discussions , and the conclusion is given in Sec.IV.\\

\noindent {\bf II. MODEL AND FORMALISM} \\

   Because of the crystalline field effect, the 3d energy level of the Mn ion
in La$_{1-x}$R$_{x}$MnO$_{3}$ and Nd$_{1-x}$R$_{x}$MnO$_{3}$ compounds is
split into a low-energy triplet( t$_{2g}$ ) and a high-energy doublet(e$_{g}$),
therefore three d-electrons of Mn ions will fill the low d- t$_{2g}$ band
with same spin, and the extra d-electrons in  Mn$^{+3}$ will fill in to
the higher d-e$_{g}$ band. These two bands are separated
about 1.5 $eV$ [15]. The three electrons in the filled d-t$_{2g}$ band
form a localized core spin of S=$\frac{3}{2}$ by Hund's rules [9,11,12],
these core spins tend to align parallelly through the double exchange
interaction between Mn$^{+3}$ ions and Mn$^{+4}$ ions [9] and can be
considered as a localized ferromagnetic background. The electrons
in d-e$_{g}$ band are mobile by hopping between Mn$^{+3}$ and Mn$^{+4}$
ions as an itinerant electron which is responsible  for
electric conduction in these systems. The Coulomb interaction of the localized
d-electrons and the conduction d-electrons can be attributed to a s-d like
exchange coupling. In this model, the Hamiltonian of this system
is  described as:
\begin{equation}
       H=H_{0}+V
\end{equation}
\begin{equation}
   H_{0}=\sum_{k\sigma} (\epsilon_{k}-\sigma \mu_{B}B) c^{\dag}_{k \sigma}
         c_{k \sigma} -\sum_{<ij>} {\it A} {\bf S}_{i} \cdot {\bf S}_{j}-
         \sum_{i}g\mu_{B}BS^{z}_{i}
\end{equation}
\begin{equation}
     V = -\frac{J}{N} \sum_{ikq} \sum_{\mu \nu} e^{i{\bf q} {\bf R}_{i}}
       {\bf S}_{i} \cdot c^{\dag}_{k+q \mu} {\bf \sigma}_{\mu \nu} c_{k\nu}
\end{equation}
where H$_{0}$ describes the bare energies of the mobile d-electrons and the
ferromagnetic background, V is the interaction between the mobile
electrons and the localized core spins S$_{i}$ on ith site. In Eq.(2),
$\epsilon_{k}$ is the energy spectrum of mobile electrons with respect
to the Fermi energy E$_{F}$, {\it A} is the effective ferromagnetic exchange
constant between manganese ions, and only the nearest-neighbor interaction
is considered. -g$\mu_{B}$B is the Zeemann energy in the magnetic field
{\bf B}. In Eq.(3)
the mobile electron is scattered from state k$\nu$ to state k+q$\mu$
by the localized spin ${\bf S_{i}}$; J is the coupling constant
 between the mobile electrons and the core spins.
We notice that in the external magnetic field and in the internal molecular
field of ferromagnetically ordered state, the mobile (or conduction) band
of the system will be split. This splitting will move the position of the
conduction band with respect to the Fermi surface, therefore the mean-field
spectrum of the conduction electron with state ${\bf k\sigma}$ is
$\epsilon_{k\sigma}=\epsilon_{k}-\sigma (\mu_{B}B+ J<S^{z}>)$.

  The scattering rate of the conduction electrons scattered by the localized
spins is:
\begin{equation}
    \omega =\frac{2 \pi}{h} \sum_{m} |<f|V|m>|^{2}  \delta (E_{f}-E_{m})
\end{equation}
where $|f>$ refers the final equilibrium state and $|m>$ denotes the
possible intermediate state of the system during the scattering process. Since
the conduction electrons are scattered into various intermediate states,
summing over all the intermediate states of the system at temperature
T, one can obtain:
\begin{equation}
    \omega = \frac{\pi}{h} \frac{J^{2}D(0)}{4} \sum_{kq\sigma}
             f_{k\sigma}(1-f_{k\sigma})f_{k+q\bar{\sigma}}(1-
             f_{k+q\bar{\sigma}})[<S^{-}_{q}S^{+}_{-q}>+
\end{equation}
\[   <S^{+}_{q}S^{-}_{-q}>+8 <S^{z}_{q}S^{z}_{-q}>].  \]
Where D(0) is the density of states of the conduction electrons near the
Fermi surface, Dirac-Fermi distribution function f$_{k}$ is
\( f_{k}=1/[e^{\beta (\epsilon_{k}-\epsilon_{F})}+1] \). As pointed out above,
the magnetic interaction dominates the properties in thses systems,
the resistivity from other scattering might be neglected. The lifetime of the
conduction electrons between two scattering is $\tau$=$\omega^{-1}$, therefore
the resistivity is given by the Drude formula,
\begin{equation}
  \rho=\frac{m^{*}}{ne^{2}\tau} = \frac{m^{*}\omega}{ne^{2}}.
\end{equation}
where n is the carrier density, m$^{*}$ is the effective mass of carriers.
Accordingly, for certain doped concentration, the effective mass and
the concentration of carriers are almost fixed, hence the resistivity can be
determined by the scattering rate of the conduction electrons. One can obtain
the temperature-dependent and field-dependent resistivity $\rho$(T,B) as
\begin{equation}
    \rho (T,B)=  \sum_{q} F(q,T,B) [<S^{-}_{q}
S^{+}_{-q}>+<S^{+}_{q}S^{-}_{-q}>+8 <S^{z}_{q}S^{z}_{-q}>]
\end{equation}
where the temperature factor F(q,T,B) is defined as:
\begin{equation}
   F(q,T,B) =\frac{\pi D(0)m^{*}J^{2}}{4hne^{2}} \sum_{k\sigma}
     f_{k\sigma}(1-f_{k\sigma})f_{k+q \bar{\sigma}}(1-f_{k+q \bar{\sigma}}).
\end{equation}
The resistivity mainly depends on the temperature factor F(q,T,B) and the
spin-spin correlation functions (longitudinal and transverse),
$<S^{z}_{q}S^{z}_{-q}>$ and $<S^{+}_{q}S^{-}_{-q}>$. When
T $<<$ T$_{c}$, or $<S_{z}> \approx $ S, the long-range magnetic order exists,
both the transverse and the longitudinal correlations have the same order of
magnitude contributions to the resistivity.
The transverse spin-spin correlation functions are chosen from Ref. [16]:
\begin{equation}
   <S^{+}_{q}S^{-}_{-q}>= \frac{2 < S^{z} >}{exp(\beta \omega_{q})-1}
\end{equation}
where $\omega_{q}$=g$\mu_{B}+zA(1-\gamma_{q}$) is the ferromagnetic spin
wave spectrum, and the longitudinal spin-spin correlation function
 $\ <S^{z}_{q}S^{z}_{-q}\ > $ is obtained by the Green's function
technique and it may be rather complicated when T$\ <$ T$_{c}$.
When T $>>$ T$_{c}$,
or $<S_{z}> \approx $ 0, the long-range magnetic order doesn't exist, and
only the longitudinal correlation $<S^{z}S^{z}>$ plays a role [17],
\begin{equation}
   <S^{z}_{q}S^{z}_{-q}>= \frac{C T}{T-T_{c}-T_{c} (\gamma_{q}-1)}
\end{equation}
where C is the Curie constant and $\gamma_{q}$ the structure factor. when
T $\approx$ T$_{c}$, the spin-spin  correlation function near the
transition temperature is also chosen from Ref.[16].

   For the case of steady-state, we can easily derive the Hall conductivity as:
\begin{equation}
       \sigma_{H}=\frac{\sigma^{2}}{ne} B_{eff},
\end{equation}
or the Hall resistivity as
\[    \rho_{H}=\frac{ne}{B_{eff}} \rho^{2},   ~~~~~~~~~~~~~~~~~~~~~~~~~~(11') \]
where B$_{eff}$ is an effective field,
\[  B_{eff}= | {\bf B}+zA\ <{\bf S} \ >/\mu_{B}|.  \]

  From Eqs.(7) and (8) we can qualitatively understand the
temperature-dependence of the resistivity. Since
\[  F(q,T,B) \propto \sum_{k\sigma}
f_{k\sigma}(1-f_{k\sigma})f_{k+q \bar{\sigma}}(1-f_{k+q \bar{\sigma}})  \]
At zero-temperature limit (T $\rightarrow$ 0 K), F(q,T,B) approaches
zero because of the Pauli exclusion principle, so $\rho(T \rightarrow 0)$
$ \approx 0$. However, at high temperature limit (T$\ >> T_{c}, E_{F}$),
F(q,T,B) $ \approx const$, the resistivity is entirely determined by the
spin-spin correlation functions, and the correlation functions are
independent of the magnetic field. In this case the
resistivities under different  applied magnetic
fields should approach to the same value. In the mediated temperature range
(0$<$k$_{B}$T  $<<$ E$_{F}$), F(q,T, B) doesn't vanish and the spin-spin
correlation functions are also finite, then the resistivity from the
spin-correlation-dependent scattering is not zero. Especially, at the
temperature near the Curie point (k$_{B}$T $\approx T_{c}$ $ <<$ E$_{F}$),
the temperature factor F(q,T,B) is smooth, the spin-spin correlation functions
have maxima, therefore the resistivity exhibits a maximum at
the transition temperature when the magnetic field is absent.\\

\noindent {\bf III. RESULTS and DISCUSSIONS } \\

   We have evaluated the resistivity $\rho$(T ,B) according to
Eqs.(7) and (8) to  compare with the experimental data.
We should point out that in order to carry out a quantitative calculation,
we need the exact forms of the transverse and the longitudinal spin-spin
correlation functions over all the temperature region and under any magnetic
field. However, such exact correlation functions are not available for two and
three dimensional Heisenberg ferromagnets since the exact solution of the
Heisenberg ferromagnet hasn't yet been obtained [16]. In this work
the approximate forms of the longitudinal and the transverse correlation
functions in Eq.(9) (T$<$ T$_{c}$) and the longitudinal correlation
function in Eq.(10) (T$>$ T$_{c}$) are adopted for our calculations.
In quasi-two-dimensional systems, the lattice constant is a=3.89 $\AA$,
and the coordinate number is z=4. Although the theoretical parameters is chosen
from La-Ca-Mn-O system only, the results may be applied in other
similar systems.

   The temperature-dependence of the spontaneous magnetization in different
magnetic fields is calculated self-consistently and shown in Fig.1.
The ferromagnet-paramagnet transition occurs within a wide temperature
range in Fig.1. Because of the critical fluctuation and
the low-dimensional character in z-direction, the
spontaneous magnetization has a long tail, the ferromagnet-paramagnet
transition is broaded. Although
the exchange integral is the same as that in the bulk, the
Curie temperatures of these quasi-two dimensional systems decrease
significantly.

  The temperature-dependence of the diagonal resistivity in La-Ca-Mn-O systems
is shown in Fig.2. The resistivity in different magnetic fields exhibits
maximum within the magnetic transition region, where the spin-correlation
scattering rate $\omega$ is very large. With the increase of the magnetic
field, the resistivity peak is driven to higher temperature and the maximum
value is reduced. This can be explained as the following. A strong field aligns
all spins in the systems and suppresses the correlation fluctuation of spins
at different sites, thus reduces the spin correlation scattering between
the conduction electrons and the local spins. Because the applied field
stabilizes the magnetic order and to destroy the magnetic order needs much high temperature
(about Zeemann energy -g$\mu_{B}$B), the maximum of the correlation function
are be moved to higher temperature region in a strong applied magnetic field.
These unusual transport behaviors due to the spin-spin correlation
fluctuation scattering agree with the recent experiment [7].
The resistivity maximum in absence of magnetic field appears exactly at the
Curie point.

   As pointed out in Sec.II, because of the thermal effect, the
high-temperature resistivity should
tend to the same value with respect to different magnetic field, however
this cannot been seen clearly in Fig.2.
It is believed that this is due to the approximation we have used in
the calculations. In high temperature region, the thermal fluctuation
is strong enough to destroy the spin alignment due to the applied magnetic
field and the exchange interaction, so the effect of the magnetic field
is smaller over the high temperature region.

    The dependence of the Hall resistivity on temperature is shown in
Fig.3. The electrons moving in the transverse direction are also scattered
by the spin-spin correlation fluctuation near the critical point,
therefore the Hall resistivity also exhibits a maximum. In the meantime,
the magnetic field {\bf B} affects the cyclotron movement of the conduction
electrons, the Hall resistivity decreases with increasing of {\bf B}, so
the maximum of the Hall resistivity will not appear at the same position
as that of the diagonal resistivity, the position of the maximum moves to
lower temperature. Since the transverse movement of the conduction
electrons is affected both by the magnetic field and by the spin-correlation
scattering, the change of the Hall resistivity may be much larger than that of
the diagonal resistivity with the variation of the magnetic field.

   The field-dependence of the diagonal and the Hall resistivities at
different temperature are shown in Fig.4 and Fig.5 respectively. We can see
that the diagonal and the Hall resistivities decrease monotonously when
the magnetic field increases. The results of the diagonal resistivity agree
with the experiments [1-4].
 We can see from Fig.4 that the diagonal resistivity ratio $\Delta R_{B}/R_{B}$
at field B=6 T can be as high as 1000$\%$. The descent resistivity with
 the increase of magnetic field is caused by  the suppress
of the spin-spin correlation fluctuation scattering .

  The resistivity due to the ${\it s-d}$ exchange interaction expressed in
Eq.(3) was studied by a number of researchers, such  as Fisher [18],
 Kubo and Ohata [19], Searle and Wang [20], and Furukawa [12]. Fisher's
work does not consider the finite size character and the
spin-splitting of the conduction electrons from the internal molecular
field and the external magnetic field, which may be important for the
MR effect in ferromagnetically ordered state. The others' studies disregard
 the spin-spin correlation functions and contain only the resistivity from
spin-flip scattering. The present work has shown that the
spin-spin correlation scattering is essential and should be responsible for
the CMR. The effect due to the spin-spin correlation and the
spin fluctuation is considerable large near the Curie point.
These theoretical results agrees with the experimental data very
well [1-4,13,14].

   Another interesting effect is the pressure-dependence of the CMR in
these systems, which is reported recently [21, 22].
The pressure has the similar influence on the resistivity as the
magnetic field does, this can be explained qualitatively within
the present model. When the pressure is increased, the exchange integral
between Mn ions also increases, correspondingly, the Curie temperature
T$_{c}$ raises and the spin correlation fluctuation is suppressed,
the resistivity maximum will move to high temperature region due to
the presence of pressure. Therefore the pressure has similar effect
on the resistivity as the magnetic filed does.

\noindent {\bf IV. CONCLUSION}\\

  From the above discussions, we can see that, (a) If the
measured temperature is far below or above the critical point T$_{c}$,
the MR $\Delta R_{B}$ is smaller. (b) If the critical transition range
is very wide, a CMR can be observed within a wide temperature region.
These observations are very important in applications. It suggests that
if we want to find or explore a new material which can be used at
the room temperature for magnetic recording and sensitive detection,
the critical point of this material should be near the room temperature,
and it should be low dimensional system, such as thin film.
Especially, for the magnetic/nonmagnetic multilayers, if the MR
is measured near its transition temperature, it would be expected a much higher MR
ratio than that observed currently. (c) the temperature dependence
of the Hall resistivity will exhibit a maximum below the peak
position of the  diagonal resistivity. Further experiments is
desired to verify these predictions.

   In conclusion, we have calculated the transport properties, such as the
diagonal and the Hall resistivity, and magnetic properties of the ferromagnetic
manganese-oxide compounds, which are in agreement with the experimental data.
The the CMR in La-R-Mn-O and Nd-R-Mn-O compounds
near the phase transition point can be attributed to the spin-spin correlation
scattering between the conduction electrons and the localized spins. The CMR
effect in thin films is enhanced by the low dimensional effect.    \\

   Acknowledgement: Authors thanks the discussion with D. Y. Xing and the
comments of Zheng-Zhong Li in Nanjing University. This work is financially
supported by the Grant of the National Science Foundations of China and the
Grant No. LWTZ-1289 of Chinese Academy of Sciences.

\newpage
\begin{center}
REFERENCES
\end{center}
\begin{enumerate}
\item R. M. Kusters, J. Singleton, D. A. Keen, R. McGreevy and W. Hayes,
       {\it Physica} {\bf B155} 362 (1989).
\item R. Von Helmolt, J. Wecker, B. Holzapfeil, L. Schultz and K. Samwer
       {\it Phys. Rev. Lett.} {\bf 71} 2331 (1993).
\item S. Jin, T. H. Tiefel, M. McCormack, R. A. Fastnacht, R. Ramesh, and
      L, H, Chen  {\it Science}. {\bf 264} 413 (1994).
\item G. C. Xiong, Q. Li, H. L. Ju, S. N. Mao, L. Senpati, X. X. Xi, R. L.
      Greene, and T. Venkatesan, {\it Appl. Phys. Lett} {\bf 66}, 1427 (1995).
\item S. Jin, H. M. O'Bryan, T. H. Tiefel, M. McCormack, and W. W. Rhodos
        {\it Appl. Phys. Lett} {\bf 66}, 382 (1995).
\item H. L. Ju, C. Kwon, Qi Li, L.Greene and T. Vankateson,
       {\it Appl. Phys. lett.} {\bf 65}, 2106 (1994).
\item P. Schiffer, A. P. Ramirez, W. Bao and S-W. Cheong,
       {\it Phys. Rev.Lett.} {\bf 75}, 3336 (1995).
\item A. Chainani, M. Mathew and D.D. Sarma
       {\it Phys. Rev.} {\bf B47}, 15397 (1993).
\item C. Zener, {\it Phys. Rev.}, {\bf 81}, 440 (1951); {\bf 82}, 403
      (1951).
\item J. Inoue and S. Maekawa, {\it Phys. Rev. Lett} {\bf 74}, 3407 (1995).
\item A. J. Millis , P. B. Littlewood, and B. I. Shraiman, {\it Phys.
      Rev. Lett} {\bf 74}, 3407 (1995).
\item N. Furukawa, {\bf J. Phys. Soc Jpn}, {\bf 63}, 3214 (1994).
\item R. Von Helmolt, J. Wecker, and K. Samwer
       {\it J. Appl. Phys.} {\bf 76}, 6925 (1994).
\item S. Jin, M. McCormack, T. H. Tiefel, R. Ramesh,
       {\it J. Appl. Phys.} {\bf 76}, 6929 (1994).
\item J. M. D. Coey, M. Viret and L. Ranno, {\it Phys. Rev. Lett.},
      {\bf 75}, 3910 (1995).
\item R. A. Tahir-Kheli, {\it Phase Transition and Critical Phenomena},
      {\bf Vol.5B}, Chapt.1 and 4, ed. by C. Domb and M. S. Green,
      (Academic Press, 1975), and some references therein.
\item R. M. White, {\it Quantum Theory of Magnetism.}  (Springer-Verlag
      series), (McGraw-Hill Inc., New York, 1970).
\item M. E. Fisher and J. S. Langer,  {\it Phys. Rev. Lett.} {\bf 20},
      665 (1968).
\item K. Kubo and N. Ohata,  {\it J. Phys. Soc. Jpn.} {\bf 33}, 21 (1972)
\item C. W. Searle and S. T. Wang,  {\it Can. J. Phy.} {\bf 48},2023 (1970)
\item Y. Moritomo, A. Asamitsu, and Y. Tokuya, {\it Phys. Rev. B} {\bf 51},
      16491 (1995).
\item Z. Arnold, K. Kamenev, M. R. Ibarra, P. A. Algarabel, C. MArquina, J.
      Blasco and J. Garcia, {Appl. Rev. Lett} {\bf 67}, 2875 (1995).

\end{enumerate}

\newpage
\begin{center}
Figures Captions
\end{center}

\noindent  Fig.1 . Temperature-dependence of the spontaneous
magnetization in different magnetic fields for La-Ca-Mn-O systems.
Parameters: {\it A}=245 K, J=500 K. (1) B=1.0 T, (2) B=15 T.
\vspace{1cm}

\noindent  Fig. 2.  Temperature-dependence of the diagonal
resistivity in different magnetic fields for La-Ca-Mn-O systems.
Theoretical parameters: {\it A}=245 K, J=500 K. (1) B=1.0 T, (2) B=15 T
\vspace{1cm}

\noindent  Fig. 3.  Dependence of the Hall resistivity
on the temperature in different magnetic fields. Theoretical parameters:
{\it A}=245 K, J=500 K. (1) B=1 T, (2) B=15 T
\vspace{1cm}

\noindent  Fig. 4.  Field-dependence of the diagonal resistivity for
La-Ca-Mn-O systems. MR ratio is 1000$\%$ for B=6 T. Theoretical
parameters: {\it A}=245 K, J=500 K, (1) T=50 K, (2) T=100 K.
\vspace{1cm}

\noindent  Fig. 5.  Dependence of the Hall resistivity on the magnetic
field for La-Ca-Mn-O systems. Theoretical parameters: {\it A}=245 K,
J=500 K, (1) T=50 K, (2) T=100 K.
\vspace{1cm}

\end{document}